# Encouraging Students' Responsible Use of GenAI in Software Engineering Education: An Explanatory Model and Two Institutional Applications

Vahid Garousi* [1, 2], Zafar Jafarov[2], Aytan Mövsümova[2], Atif Namazov[2], Hüseyn Mirzəyev[3]

1: Queen's University Belfast, United Kingdom
2: Azerbaijan Technical University, Azerbaijan
3: Karabakh University, Azerbaijan

v.garousi@qub.ac.uk, {vahid.garousi, zafar.cafarov, aytan.movsumova, atif.namazov}@aztu.edu.az, huseyn.mirza@karabakh.edu.az

**Abstract:**

*Background:*

Generative AI (GenAI) tools such as ChatGPT and GitHub Copilot are increasingly used in Software Engineering Education (SEEd), raising concerns and risks about shallow understanding, reduced critical thinking, academic integrity risks, and long-term skill development of students.

*Objective:*

To address the above risks, we have designed and present in this paper a pragmatic explanatory model that links professor actions, student factors, and GenAI tool characteristics to observable educational outcomes.

*Approach:*

The model was developed using design-based research (DBR) and will be applied in two SEEd contexts: (1) redesigning four lab assignments in a final-year Software Testing course, and (2) embedding GenAI competencies across the curriculum of a Software Engineering (SE) BSc program. Interventions included GenAI usage declarations, validation tasks, peer-review of AI outputs, and career-oriented messaging.

***Results:***

**TBD**

***Conclusions:***

**TBD**

* Corresponding author

# Table of Contents

## 1 INTRODUCTION

The emergence of generative AI (GenAI) tools—such as ChatGPT, GitHub Copilot, and Google Gemini continues to revolutionize every single discipline, including software engineering (SE) and also the educational counterpart of each field, e.g., Software Engineering Education (SEEd) for SE. These tools leverage large language models and generative architectures to produce human-like responses, code, and content.

In the context of SE, GenAI is reshaping how software engineers learn and conduct their work tasks [1], from AI-assisted coding to automated test-case generation. AI has also introduced transformative possibilities in the landscape of education [2-5]. In computing and SE disciplines, where problem-solving, programming (code development) are central, the adoption of GenAI tools has been particularly rapid [4-6].

In this context, it becomes essential to critically compare and contrast the role of AI in the practice of SE and its role in the education of SE (SEEd). While AI in SE practice is primarily leveraged to enhance productivity, automate tasks (e.g., test generation, code completion), and support decision-making (e.g., defect prediction), its application in SEEd can provide different sets of benefits, such as enabling personalized learning through intelligent tutoring systems or assisting students in debugging code. In both contexts (SE practice versus SEEd), responsible and proper use of AI can yield benefits: in SE, it can reduce development costs and time-to-market; in SEEd, it can support diverse learning styles and foster student engagement. However, misuse of AI in either context can lead to negative consequences: in SE, overreliance on AI-generated artifacts without human review may result in undetected defects or security vulnerabilities; in SEEd, uncritical dependence on AI tools may hinder the development of students' (learners) SE and problem-solving skills and academic integrity, such as students submitting AI-generated assignments without understanding the content. Therefore, clear guidelines and critical thinking in both contexts are crucial to ensure AI contributes constructively, and to prevent negative side effects.

SEEd, as a subfield of computing education [5, 6], is witnessing both enthusiasm and disruption in how students and instructors interact with SE coursework, assessments, and educational artifacts [7-10]. Both educators and students are actively exploring the use of GenAI tools in educational activities. For students, GenAI can provide instant support in understanding complex concepts, help them learn step by step with clear explanations, generate code snippets, debug code, and even compose entire software artifacts. These tools can serve students as intelligent assistants during self-paced learning [11]. Educators, on the other hand, benefit from GenAI through automated content creation, assignment drafting, grading support, and personalized feedback mechanisms [12]. When used responsibly, GenAI has the potential to elevate both teaching effectiveness and learner productivity.

Given the importance of this issue (AI's increasing role in education), UNESCO, as part of its United Nation's Sustainable Development Goals program, has published a recent book related to AI competency for students [11], and another similar book on the topic, specific for educators [12].

However, the integration of GenAI into the educational process is not without challenges. Concerns are mounting that students may rely on these tools not as learning aids, but as shortcuts—bypassing the cognitive effort required for deep understanding [9]. Such behavior is risky and could leading to superficial learning, undermine the development of core competencies such as problem solving and code validation, and raises academic integrity concerns [13]. Educators now face a new pedagogical dilemma: how to encourage students to harness the potential of GenAI while preventing its misuse. Thus, the need for structured, discipline-aware strategies to guide responsible GenAI use in all fields of education, including SEEd, is becoming increasingly crucial.

In SEEd, students are expected to synthesize SE knowledge, apply SE techniques (e.g., software testing) in project settings, and make professional judgments. In SEEd contexts, overreliance of students on GenAI could inhibit the development of critical skills such as algorithm design, defect detection, and software peer review. These concerns are especially pressing in hiring new SE graduates, where employers increasingly expect not just fluency in prompt engineering [14], but critical thinking in all aspect of SE and the ability to validate AI-generated outputs, e.g., code, test cases. Students who bypass cognitive effort with GenAI may underperform in job interviews, and later real-world SE tasks [15, 16].

Therefore, educators in all fields of education in general, and SE education in particular, need to take proper actions to encourage students' responsible use of GenAI in their learning and educational activities. To address this need, we introduce in this paper a structured, pragmatic and practical approach. We first identify the key factors influencing students' use of GenAI tools, and group them into three categories: educator (professor)-related, student-related, and tool-related factors. We then design an explanatory (conceptual) model that maps the factors to a number of outcome factors, e.g., developing critical thinking, and learning to validate and verify AI-generated outputs. Bu following design-based intervention guidelines [17], the model identifies intervention points such as assignment design, and including peer validation of AI artifacts in coursework. To assess the applicability and usefulness of the model, we have empirically applied the model and approach in two different real educational settings: (1) a Software Testing course, and (2) the curriculum design of a new SE program.

This work addresses a key gap in the GenAI-in-education landscape: the lack of structured, theory-informed guidance for designing interventions that go beyond rule-based prohibitions or ad hoc classroom decisions. In response, we propose an explanatory model that maps how instructor behaviors, student factors, and GenAI tool characteristics interact to influence learning outcomes. Our model is designed not only as an explanatory framework but also as a practical planning tool. It is grounded in educational theory, refined through design-based research (DBR) [17], and evaluated through two empirical applications in SEEd settings.

We do not advocate banning GenAI usage in education, but to foster its responsible use, since no matter what, students and software engineers will indeed continue using GenAI. The model and approach presented in this work aim to help students evolve from AI *consumers* to AI-*literate* engineers. In doing so, we contribute both a theoretically-grounded model and practical design-based interventions [17] for SE educators navigating the GenAI era.

While several elements of the explanatory model and design-based interventions proposed in this paper are tailored to SEEd—such as validation and peer review of AI-generated software artifacts (e.g., code, tests, SE documentation), the broader structure of the model remains adaptable to other fields. Many of the underlying principles, such as fostering student reflection, mitigating overreliance on GenAI, and embedding responsible usage policies, are applicable across other areas of computing education (e.g., data science, cybersecurity) and even non-computing disciplines. As such, this work contributes not only to SEEd but also offers a transferable framework for educators in various fields navigating the challenges and opportunities presented by GenAI in education.

In summary, this paper makes three main contributions: (1) We propose an explanatory model that captures factors influencing responsible GenAI use in SE education; (2) We redesign a real-world SE testing lab to align with this model and evaluate its integration; and (3) We provide practical insights into how students interact with GenAI tools and develop verification skills through iterative use and critical evaluation.

The remainder of this paper is structured as follows. Section 2 reviews the related work and pedagogical theories. Section 3 introduces the proposed explanatory model for responsible GenAI use in SE education. Section 4 outlines the empirical application of the model in two distinct educational contexts. Section 5 discusses reflections. lessons learned, implications for educators and also limitations of this work. Section 6 concludes and discusses possible future directions.

The authors shall note that in addition to citing and discussing academic peer-reviewed literature in the rest of this paper, they also cite credible grey literature (non-peer-reviewed sources) too, since the topic of this paper (using AI in education) is very actively discussed in the grey-literature sources also. We shall also remind the reader that benefitting from the value of grey literature in scholarly research has been recognized in various fields of science, e.g., SE, AI and education, e.g., [18-24]. However, the authors still explicitly note that the cited grey literature sources are to be treated as illustrative rather than evidentiary.

## 2 BACKGROUND AND RELATED WORK

The integration of GenAI tools in SE and computing education is transforming traditional pedagogical dynamics. GenAI tools such as ChatGPT and Copilot have introduced new educational opportunities, enabling students to query complex topics, generate working code, and do code reviews. While these tools can augment learning if used responsibly, they also risk undermining foundational learning objectives if treated as replacements for human effort (of students) and their cognitive engagement [13, 25].

To establish a foundation for our work, this section synthesizes relevant academic and grey literature on the use of GenAI in CS/SE education, in the following focused topics:

- GenAI's role in professional SE versus SE education (Section 2.1)
- Adoption patterns of GenAI in CS/SE education (how GenAI is used by students) (Section 2.2)
- SWOT analysis of GenAI in CS/SE education (Section 2.3)
- Relationship of GenAI with the Bloom's taxonomy (Section 2.4)
- Emerging frameworks for responsible usage of AI in education (Section 2.5)
- Emergence of institutional policies on acceptable usage of GenAI by students and instructors (Section 2.6)
- Summarizing contributions and novelty of this work (Section 2.7)

### 2.1 GenAI in professional SE vs. SE education

The application of GenAI in professional SE differs distinctly from its use in educational settings. In the SE industry, GenAI tools are primarily employed to enhance productivity, automate routine coding tasks, and assist in software design and testing. Engineers utilize these tools to streamline workflows and reduce development time [26].

Conversely, in educational contexts, the focus is on using GenAI to support learning objectives, develop problem-solving skills, and foster a deeper understanding of SE principles. The pedagogical goal is not merely to produce code but to cultivate the analytical and critical thinking abilities essential for professional practice [27].

This dichotomy underscores the need for educational strategies that balance the practical benefits of GenAI with the imperative to develop foundational competencies. Educators must design curricula that integrate GenAI in ways that enhance learning without compromising the development of essential skills.

### 2.2 Adoption patterns of GenAI in CS/SE education (how GenAI is used by students and educators)

The adoption of GenAI tools in computing and SE education has been rapid and multifaceted. Students now routinely employ tools such as ChatGPT, GitHub Copilot, and Google Gemini across a range of learning activities, including code generation, debugging, synthesizing documentation, test-case writing, and even explanation of theoretical SE concepts. These tools act as on-demand assistants, especially valued by students during self-paced assignments and outside standard instructional hours.

For instance, a survey study [26] conducted at Utrecht University, Netherlands, revealed that students frequently use GenAI to assist in programming tasks and to clarify complex SE topics. Other empirical studies [6, 27] echo this finding, noting that students use GenAI not only to solve tasks but also to compare outputs, explore alternative implementations, and scaffold their understanding when course materials seem too abstract.

Educators, meanwhile, are grappling with how best to integrate GenAI into curricula [28]. Some instructors have begun designing assignments that explicitly allow or require the use of GenAI, accompanied by validation, critique, or peer-review tasks to ensure deep engagement. Others remain cautious, citing concerns over academic dishonesty, learning erosion, and the lack of explainability in AI-generated outputs.

Despite its benefits, GenAI usage patterns reveal a spectrum of behaviors. Some students use it responsibly as a learning scaffold, while others rely on it to bypass cognitive effort—copy-pasting unverified code or reports. This variance underscores the urgent need for structured pedagogical interventions that help students develop both AI literacy and the ability to evaluate and adapt AI-generated content within CS and SE learning contexts.

### 2.3 SWOT analysis of GenAI in CS/SE education

The integration of GenAI tools into CS and SE education has sparked widespread interest among educators and researchers. A useful way to analyze the pedagogical implications of GenAI is through the SWOT framework—examining its strengths, weaknesses, opportunities, and threats in academic contexts [2, 5, 27, 29, 30].

**Strengths:** Many studies have studied strengths (benefits) of GenAI in education. Among its strengths, GenAI is frequently praised for its ability to provide personalized, real-time support to students [29, 30]. By helping with syntax correction, explanation of algorithms, generating tests, and many other use cases, tools like ChatGPT can reduce the cognitive burden on novice learners and enhance learning efficiency [27, 30]. GenAI also offers ubiquitous access to support, enabling students to receive assistance beyond traditional university sessions (e.g., classroom and open office hours). From an inclusion perspective, these tools may help level the playing field for students with limited access to human tutoring or for those learning in a second language [2, 27, 29].

**Weaknesses:** The above strengths are accompanied by considerable weaknesses. One of the most cited concerns is over-overreliance of students on GenAI, rather than being self-dependent (solving all the coursework on their own). Students may become dependent on AI-generated outputs without developing a deep understanding of the underlying principles. This dependency can erode critical thinking, particularly in higher-order SE tasks such as architecture evaluation or debugging complex systems [2, 27, 30]. Moreover, GenAI tools are known to "hallucinate" incorrect or misleading content. Without sufficient AI literacy, students may struggle to evaluate the validity of generated outputs. The black-box nature of many GenAI tools further complicates their educational use, especially in contexts requiring explainability and traceability.

**Opportunities:** Despite the weaknesses, GenAI creates significant opportunities for curricular transformation. Educators are beginning to embed GenAI literacy, prompt engineering [14], and AI-critique exercises into computing curricula [27, 30]. When guided by intentional instructional design, GenAI can help foster metacognition (thinking about thinking), enable adaptive learning environments, and expose students to modern SE practices such as automated testing and continuous integration pipelines. In this context, metacognition refers to students' awareness and control of their own thinking and learning processes, while using GenAI. When GenAI is used with intentional instructional design, it can prompt students to: reflect on how they solve problems, and monitor their understanding of SE concepts. These opportunities are particularly relevant as GenAI becomes embedded in industry workflows, raising the importance of preparing AI-aware graduates [2, 5].

**Threats:** At the same time, several threats must be acknowledged. Academic integrity remains the most visible concern, as students may use GenAI to generate code, reports, or assignments without acknowledgment or understanding. Institutions worldwide are revisiting their plagiarism and misconduct policies considering these new capabilities of AI tools [5, 30].

Other concerns related to AI ethics, such as bias, data privacy, and intellectual property, are becoming increasingly relevant and must be integrated into responsible SE education practices [27, 29].

We shall clarify that, while weaknesses refer to internal limitations within the educational process—such as students' overreliance on GenAI, diminished critical thinking, or inadequate AI literacy—threats denote broader external risks that challenge the integrity and future of SE education. These include rising academic misconduct facilitated by GenAI, institutional failure to keep curricula aligned with fast-evolving AI technologies, and unresolved ethical issues such as bias, data privacy, and intellectual property. Addressing weaknesses typically involves pedagogical and course-level interventions, whereas mitigating threats requires systemic responses at the institutional, policy, or societal level.

In summary, the above brief SWOT analysis, synthesized from the related work [2, 5, 27, 29, 30], highlight the dual nature of GenAI's impact on SE and computing education. Harnessing its benefits requires proactive design, policy innovation, and instructional scaffolding that promotes responsible, reflective, and ethically informed student engagement with AI tools.

## 2.4 Relationship of GenAI with the Bloom's taxonomy

Bloom's taxonomy was proposed in 1956 [31] and has long served as a popular framework for categorizing educational goals, emphasizing the progression from lower-order to higher-order cognitive skills. The introduction of GenAI tools presents both opportunities and challenges in this context. Several recent AI-aware revisions (extensions) of the Bloom's taxonomy has been proposed recently, e.g., [32], as shown in Figure 1.

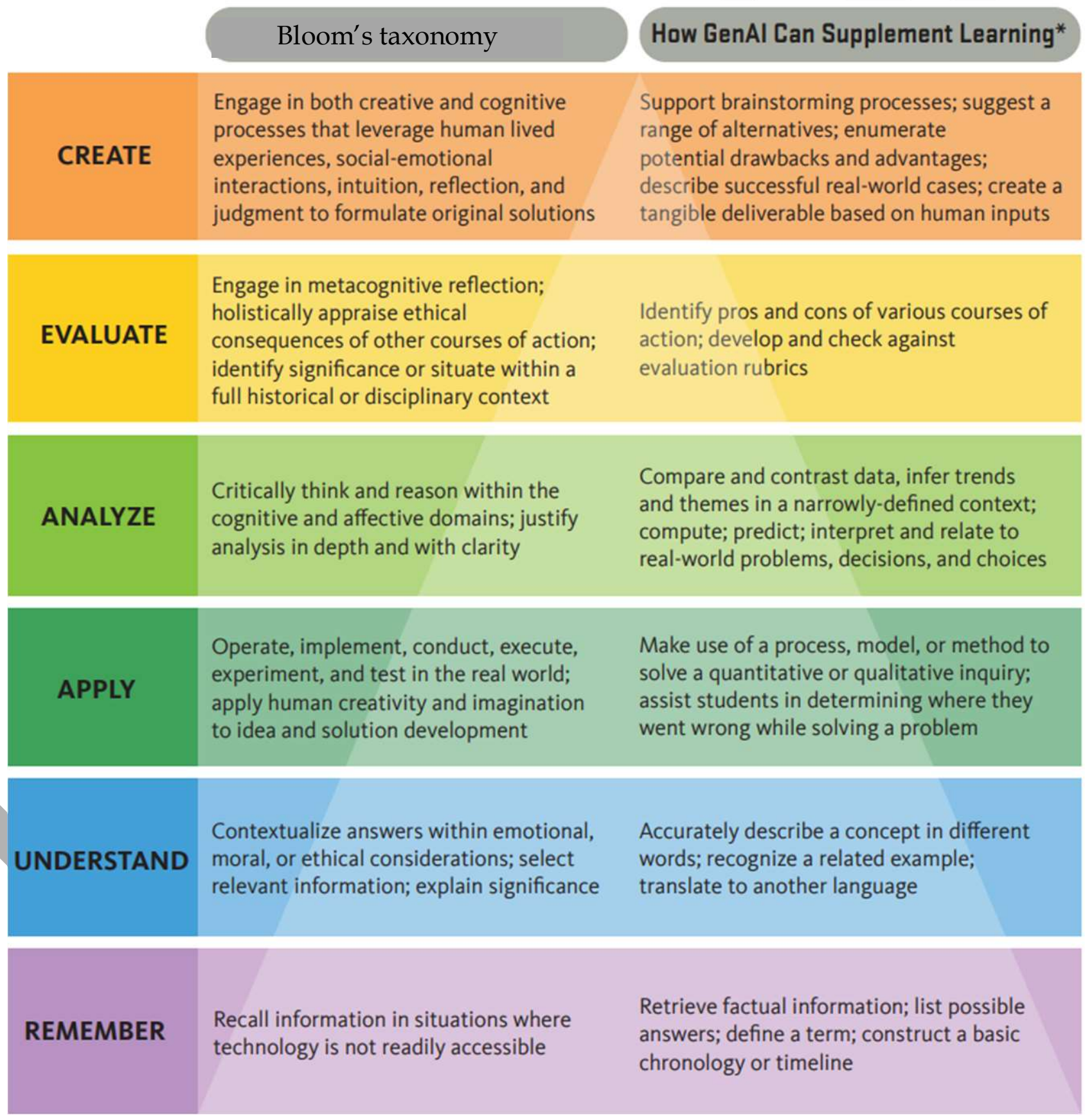

**Figure 1- Bloom's taxonomy [31], and how GenAI can supplement learning in each level (from [32])**

GenAI can effectively support the lower-order cognitive levels of Bloom's taxonomy—such as remembering and understanding—by providing instant access to information, definitions, and conceptual explanations. However, its capacity to support higher-order skills like analysis, evaluation, and creation is more challenging and less straightforward. While some educators believe GenAI can enhance these skills by offering diverse perspectives and aiding ideation, others caution that uncritical or excessive reliance on GenAI may hinder the development of deep thinking, creativity, and independent problem-solving [33], an opinion which the authors of this paper also agree with.

To address these concerns, pedagogical strategies are being explored to integrate GenAI in ways that promote active learning and critical engagement of learners (students). For example, studies have reported that assignments requiring students to critique or improve AI-generated content can encourage deeper cognitive processing and reinforce higher-order skills [33].

## 2.5 Frameworks for responsible use of GenAI in education

The responsible integration of GenAI into education necessitates the development of frameworks that address ethical, and pedagogical considerations. Many educational institutions and organizations have proposed guidelines to ensure proper use of AI in educational contexts.

The World Economic Forum (WEF), for instance, has outlined principles [34, 35] emphasizing transparency, accountability, and inclusivity in AI deployment within schools. Similarly, UNESCO has developed competency frameworks aimed at equipping students [11] and educators [12] with the skills to engage with AI responsibly.

The UNESCO AI competency framework for students [11] articulates competencies to prepare students as responsible users and co-creators of AI. It highlights fostering a critical understanding of AI's ethical, social, and environmental impacts, encouraging students to prioritize human agency, uphold transparency, and evaluate when AI tools should or should not be used. The framework promotes embedding ethical judgment and civic responsibility in AI interactions to support sustainable, inclusive societies. Both frameworks reinforce that responsible GenAI use demands not only technical literacy but also strong ethical and humanistic foundations.

Similarly, the UNESCO AI competency framework for teachers [12] emphasizes empowering teachers with the knowledge, skills, and ethical principles needed to navigate the responsible integration of AI in education. It stresses a human-centered mindset, the ethics of AI, and teacher accountability, warning against overreliance on AI technologies that could undermine critical thinking and human agency. Responsible use is framed around ensuring inclusivity, protecting student data privacy, promoting sustainability, and maintaining human oversight over AI-driven decision-making.

Three other frameworks in this area are: the PAIR (Problem, AI, Interaction, Reflection) framework [36], the AI-ICE model (Interaction, Cognition, Ethics) [37], and the AI-Lab framework [38]. The PAIR framework [36] suggest a four-step approach for guiding learners when using GenAI tools.: (1) define meaningful problems, (2) select suitable AI tools, (3) interact with the tools critically, and (4) reflect on the outcomes and ethical implications. Emphasizing human-centric, skill-centric, and responsibility-centric principles, the PAIR framework fosters critical thinking, ethical reasoning, creativity, collaboration, and adaptability. By integrating AI literacy into curricula, educators can prepare students to use AI tools effectively and navigate the associated ethical and societal challenges.

The AI-ICE model (Interaction, Cognition, Ethics) [37] encourages critical thinking about GenAI's role in learning environments by focusing on three dimensions: how learners interact with AI tools, how cognition is shaped by these tools, and what ethical questions arise from their use. While our model aligns with the AI-ICE framework in its emphasis on cognitive engagement and ethics, it extends that work by offering fine-grained, actionable levers for intervention (e.g., assignment design, validation prompts, peer-review strategies), particularly within technical domains such as SE.

The AI-Lab framework [38, 39] emphasize GenAI as a "thought partner" in programming education, guiding students to reflect on its outputs rather than blindly accept them. This idea resonates with our model's emphasis on validation tasks and reflective engagement. However, our explanatory model further incorporates sociotechnical interactions (e.g., peer pressure, time constraints, tool characteristics), and captures system dynamics—such as feedback loops—that many existing models do not.

## 2.6 Emergence of institutional policies and guidelines on acceptable usage of GenAI by students

At the institutional level, universities are also preparing policies and guidelines that define acceptable uses of GenAI, address concerns about the academic-integrity-related issues, and provide support for faculty and students navigating this evolving landscape. These policies are crucial for fostering an environment where GenAI can be leveraged to enhance learning while mitigating potential risks.

For example, the UK Russell Group of universities' principles on GenAI in education (its 2025 version) [40] emphasize five key issues: (1) supporting students and staff to become AI-literate; (2) equipping staff to guide responsible AI use; (3) adapting teaching and assessment methods to integrate ethical GenAI usage; (4) ensuring academic integrity is upheld; and (5) fostering collaboration across institutions. Notably, the principles stress transparency, privacy risks, bias awareness, validation of AI outputs, and the critical need for human judgment. Universities are encouraged to embed GenAI reflection into pedagogies while ensuring fairness of access and academic rigour.

As another example, the guidance by Queen's University Belfast on the use of AI in assessments (2024) [41] offers policy recommendations for responsible AI use by students. It clarifies that using GenAI without proper attribution or understanding constitutes academic misconduct. Students are required to declare AI use, and assessments are being redesigned to encourage authentic learning and critical validation of AI outputs. Staff are supported to create assignments less vulnerable to uncritical GenAI reliance, and students are advised that GenAI tools must be treated as support resources—not substitutes for independent reasoning.

Another UK university, the University of Exeter has developed adaptable guidelines to promote the ethical and responsible use of generative AI (GenAI) tools [42]. These guidelines emphasize enhancing learning outcomes without replacing core educational objectives. Transparency is mandated, requiring both staff and students to disclose when and how GenAI tools are utilized in their work. The use of GenAI must comply with data protection regulations, ensuring no sensitive information is input into non-GDPR-compliant tools. Critical engagement with GenAI outputs is encouraged, promoting evaluation of accuracy and awareness of potential biases to uphold academic integrity.

### 2.7 Summary of gaps, contributions and novelty of our work

Despite the growing body of literature exploring the role of GenAI in education, most existing studies remain broad and generic, offering high-level SWOT analyses or ethics-oriented guidelines without grounding them in specific disciplinary contexts such as SEEd. Furthermore, there is limited work that links the use (or misuse) of GenAI tools to observable educational outcomes or underlying cognitive skill development within SEEd.

A key gap we identified is the absence of a structured framework that causally maps how student interactions with GenAI tools influence their learning behaviors, cognitive development, and long-term readiness for SE practice. While Bloom's taxonomy and frameworks like PAIR provide useful lenses, they do not offer a cohesive causal explanation that can be used to inform interventions, course design, or institutional policies.

To address this, our primary contribution is the development of a explanatory model that captures key factors shaping responsible and effective use of GenAI in SEEd. This model is informed by literature on AI literacy, cognitive skill development, and GenAI capabilities, and explicitly highlights risks such as overreliance and bypassing of deep learning processes. The model is designed to be actionable, helping educators and institutions understand not only what risks exist but why they occur and how they can be addressed. To demonstrate the model's practical relevance and adaptability, we present two institutional applications of it, in Section 4.

Overall, the novelty of our work lies in (1) proposing a discipline-specific explanatory model of GenAI use in SEEd, and (2) demonstrating its real-world application across two distinct institutional contexts. To our knowledge, this is the first work to offer such an explanatory and applied approach to fostering responsible GenAI use in SEEd, which can also be adapted to other fields.

## 3 EXPLANATORY MODEL OF RESPONSIBLE GENAI USE IN SE EDUCATION

To support educators in promoting responsible and critical use of GenAI tools in SEEd, we have developed and present in Figure 2 an explanatory model that captures the key pedagogical, behavioral, and technological dynamics underlying students' usage of AI tools in their learning and educational activities.

We developed the model based on an extensive literature review, classroom observations since 2022 (when ChatGPT was first introduced), and iterative refinement through design-based interventions. The model integrates ideas from Bloom's Taxonomy [31, 33, 43], constructivist learning theory [44], and sociotechnical systems theory [45], and is structured to help instructors understand and intervene in the causes—not just the symptoms—of GenAI misuse.

We drew on constructivist learning principles [44] to reflect the idea that students actively build understanding through their interactions with GenAI, rather than passively receiving information. This perspective helps explain how certain uses of GenAI can either support or hinder learning depending on the depth of engagement. In parallel, sociotechnical systems thinking [45] guided us to consider the broader context in which GenAI is used—acknowledging that student behavior is shaped not only by personal choices but also by institutional policies, course design, and tool affordances. These perspectives allowed us to design a model that captures both individual learning pathways and systemic influences.

Rather than offering a rigid protocol, the model serves as a flexible framework. It enables educators to identify intervention points, such as course design, peer norms, or AI tool characteristics, that influence how students engage with GenAI in their coursework. The model that we have designed (Figure 2) is particularly relevant / useful in SEEd contexts, where the use of GenAI tools to automatically generate code or other artifacts is sometimes mistaken for genuine understanding—making it critical to reinforce the need for student judgment, critical thinking, and oversight. We discuss next the model's structure, the involved actors and their relationships.

To frame our approach pedagogically and when designing this model, we have utilized the Bloom's Taxonomy—both its classical version [31] and a recent AI-aware 2024 extension of it, proposed by Oregon State University [43]. Many student uses of GenAI—such as summarizing software documentation, generating code, or test cases—tend to remain at the lower levels of Bloom's Taxonomy, focused on remembering and understanding.

Without deliberate guidance by SE educators, these uses often bypass meaningful learning. To promote deeper engagement, SE educators should design tasks that require students to critically evaluate AI-generated artifacts—such as checking code correctness, crucial review of AI-written requirements, or improving GenAI-generated test cases. Such activities would help learners develop essential skills such as analysis, evaluation, and creation, ensuring they learn *with* AI rather than simply *from* it. The revised Bloom's model, a good work by Oregon State University [43], emphasizes deeper cognitive engagement, especially in the age of AI. Students must still learn to interpret, validate, and critique GenAI outputs. Educators are thus called to design assessments and coursework that emphasize human oversight, reflective learning, and AI literacy.

By aligning interventions with Bloom's higher-order objectives, the model and approach presented in this work aim to help students evolve from AI consumers to AI-literate engineers. In doing so, we contribute both a theoretically-grounded model and practical design-based interventions for SE educators navigating the GenAI era.

### 3.1 Actor groups and their influencing factors

As shown in Figure 2, the model organizes influencing factors across three actor groups: professors (educators), students, and GenAI tools. Each group exerts distinct types of influence and interacts dynamically with the others. In the model, we have included different factors for each actor group, as discussed next.

The "+/–" directional links in the model (Figure 2) were initially proposed conceptually, based on our observations and experience in our educational contexts (teaching various courses in the past ~20 years). Furthermore, we will see in Section 4 that "+/–" directional links were subsequently observed empirically in our first case study (the software testing course and student behavior while using GenAI). However, we shall clarify that not all links are empirically validated at this stage in-depth, and the model should therefore be interpreted as explanatory rather than causal.

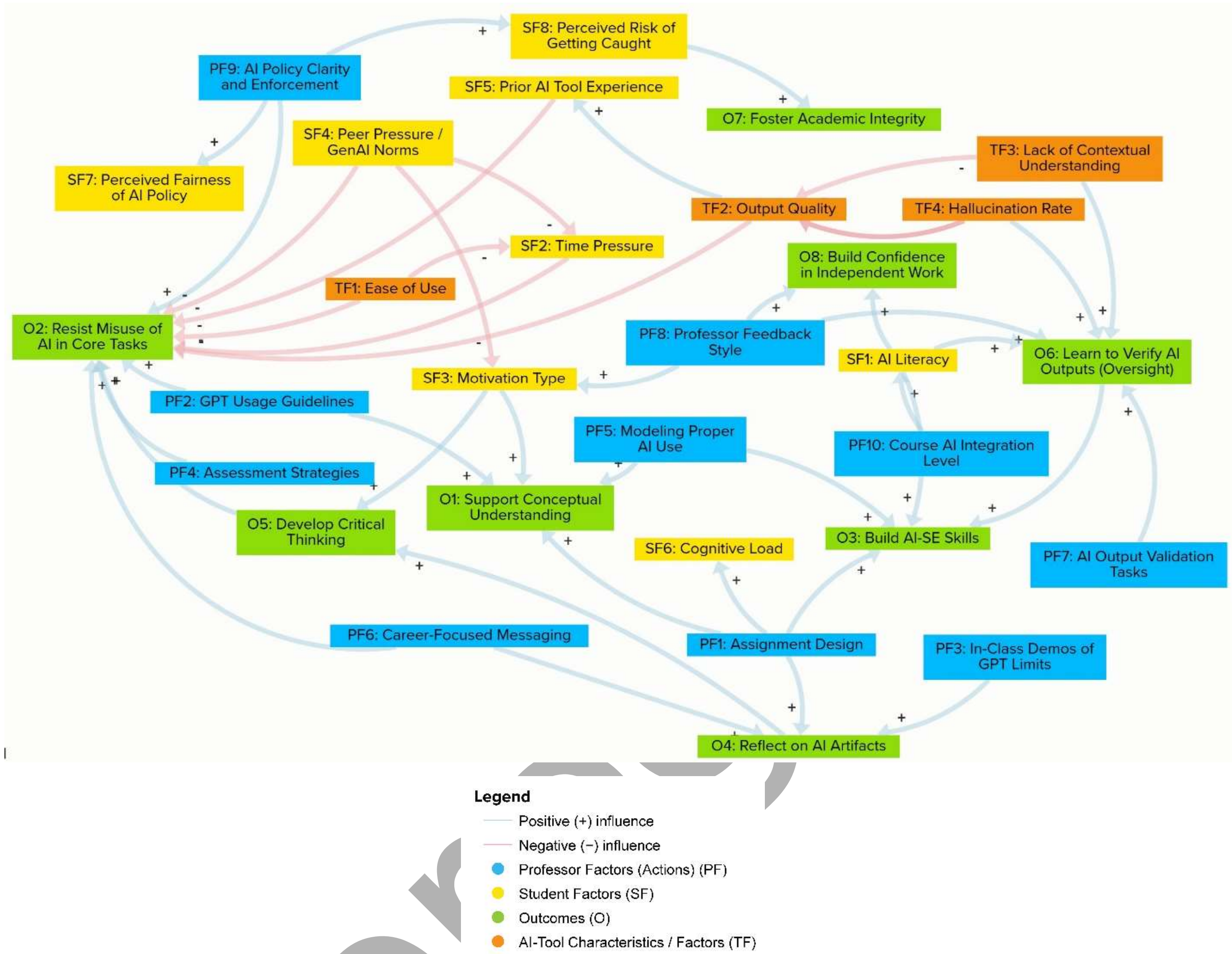

**Figure 2- Explanatory model of responsible GenAI use in SE education (generated using the kumu.io tool). A high-quality PDF file of the model and the kumu.io-compatible JSON file can be found in this DOI URL [46]**

### 3.1.1 Professor (educator) factors

Professors (educators) act as the architects of the learning environment. For instance, we have modeled professor's actions in the context of a given course using 10 factors, labeled as PF1..PF10 (PF standing for Professor Factors) [47-50].

Professors influence students' GenAI use through a range of interconnected actions. First, AI-aware assignment design (PF1) plays a foundational role—assignments that explicitly require analysis, critique, or refinement of AI outputs encourage meaningful engagement. Next, the presence of clear GenAI usage guidelines (PF2) helps students understand the boundaries of acceptable use. In-class demonstrations of GenAI's limitations (PF3)—such as live critiques of flawed AI-generated artifacts—can effectively reveal tool weaknesses and foster critical thinking. Professors' assessment strategies (PF4), including rubrics that reward thoughtful AI integration or penalize blind reliance, shape how students approach AI in their work [47, 49].

Through modeling proper AI use (PF5), instructors set examples of responsible interaction with GenAI—for instance, by showing how to refine or verify AI outputs in real time. Career-focused messaging (PF6) about industry expectations helps reinforce the importance of AI literacy and human oversight (linked to Objective O3: Build AI-SE Skills). Incorporating AI output validation tasks (PF7), such as requiring students to explain and correct GenAI-generated code, strengthens critical engagement. The professor's feedback style (PF8) is also influential—specific, formative feedback that references how students used or challenged AI promotes higher-order learning. Ensuring clarity and consistent enforcement of AI policies (PF9) reduces ambiguity and helps uphold academic integrity. Finally, the overall level of AI integration into the course (PF10) signals how central GenAI is to learning, shaping students' perceptions of its legitimacy and value [48, 49]. For easier access and reference, we provide in Table 1 the list of all factors, categorized by the four actor groups.

Together, these ten professor-driven factors represent a comprehensive and actionable set of levers through which instructors can shape students' attitudes, behaviors, and cognitive engagement with GenAI tools. Rather than treating GenAI as a passive content aid, these factors help embed it into the learning environment in ways that promote responsible use, deeper thinking, and alignment with professional SE practices. Importantly, the factors are not isolated interventions but mutually reinforcing elements that, when applied intentionally, can shift classroom culture toward critical, reflective, and skill-building uses of GenAI.

**Table 1-List of all factors, categorized by the four actor groups**

| | |
|---|---|
| **Professor factors (actions) (PF)**<br>• PF1: AI-aware Assignment Design<br>• PF2: GenAI Usage Guidelines<br>• PF3: In-Class Demos of GenAI Limits<br>• PF4: Assessment Strategies<br>• PF5: Modeling Proper AI Use<br>• PF6: Career-Focused Messaging O3: Build AI-SE Skills<br>• PF7: AI Output Validation Tasks<br>• PF8: Professor Feedback Style<br>• PF9: AI Policy Clarity and Enforcement<br>• PF10: Course AI Integration Level | **Educational Outcomes (O)**<br>• O1: Using AI's support for conceptual understanding of SE topics<br>• O2: Resisting misuse of AI in core SE tasks<br>• O3: Developing AI-assisted SE skills<br>• O4: Reflection and critical evaluation of AI artifacts (higher-order levels of the Bloom's taxonomy)<br>• O5: Developing and preserving critical thinking skills during AI-supported learning<br>• O6: Developing skills to verify AI-generated outputs (oversight)<br>• O7: Upholding academic integrity in AI-supported learning environments<br>• O8: Building confidence in judging and using AI responsibly |
| **GenAI-tool characteristics (factors) (TF)**<br>• TF1: Ease of Use<br>• TF2: Output Quality<br>• TF3: Lack of Contextual Understanding<br>• TF4: Hallucination Rate | **Student factors (SF)**<br>• SF1: AI Literacy<br>• SF2: Time Pressure<br>• SF3: Motivation Type<br>• SF4: Peer Pressure / GenAI Norms<br>• SF5: Prior AI-tool Experience<br>• SF6: Cognitive Load<br>• SF7: Perceived Fairness of AI Policy<br>• SF8: Perceived Risk of Getting Caught |

### 3.1.2 Student factors

Students are obviously a key actor as well. They play an active role in shaping how GenAI is used in SEEd contexts. Their level of AI literacy (SF1)—that is, their ability to understand the capabilities and limitations of GenAI tools—fundamentally influences whether they use these tools to support or bypass learning [30, 51].

Time pressure (SF2), such as looming deadlines or workload overload, often drives students to use GenAI as a shortcut. Their underlying motivation type (SF3) also matters: students driven by intrinsic goals (e.g., mastery) are more likely to use GenAI responsibly, whereas extrinsically motivated students may prioritize increased efficiency when using GenAI to get higher grades over understanding [4, 52].

Peer pressure and emerging GenAI norms within the classroom (SF4) also shape students' sense of what is acceptable, especially in the absence of explicit instructor messaging. A student's prior AI-tool experience (SF5) can increase confidence and skillful usage—or lead to overconfidence and misuse. In high cognitive load situations (SF6), students may rely on GenAI to reduce effort, particularly when coursework tasks are overly complex [11, 26].

The perceived fairness of the AI policy (SF7) also plays a role: if policies are unclear or seem unjust, students may disengage from them. Finally, the perceived risk of getting caught (SF8) using GenAI improperly influences how closely students follow expectations [11, 36].

Collectively, the above eight factors highlight that students are strategic, not passive; they make nuanced decisions about GenAI use based on internal goals, external pressures, and the learning environment around them.

### 3.1.3 GenAI-tool factors

GenAI tools may be seen as neutral in how students interact with them in educational settings, but their design and functionality can significantly shape how students interact with them. Our model includes four key tool-level factors (TF1–TF4) that influence student behavior and learning outcomes in SEEd.

First, the usability (ease of use) (TF1) of GenAI tools—through intuitive interfaces, prompt suggestions, and rapid responses—can encourage habitual use. Tools that are frictionless and responsive may reduce students' hesitation to rely on them, particularly under time pressure or cognitive load.

Second, the quality of outputs generated by GenAI tools (TF2) plays a central role in shaping user trust and reliance. High-quality outputs—those that are accurate, complete, and well-structured—can reinforce student confidence in using GenAI tools. For example, when GenAI produces technically sound test cases or logically coherent design diagrams, students may find these outputs genuinely helpful in understanding and completing their coursework. This factor reflects the positive potential of GenAI when outputs meet task-specific expectations and support learning objectives.

The third factor relates to the widely-discussed "context" problem in AI [53, 54]. GenAI tools often demonstrate a lack of contextual understanding (TF3). One critical limitation of current AI systems is their inability to grasp deeper semantic meaning in user prompts. While such systems excel at identifying patterns across vast datasets, they often misinterpret intent when the surrounding context is vague or complex. As Hoffman argues, this shortfall arises from AI's inability to truly "understand" data—it detects correlations, but lacks human-level interpretation skills [55].

GenAI tools may produce plausible outputs that fail to capture task-specific requirements or domain constraints. In SEEd, this can result in code that compiles but misses edge cases, or documentation that lacks semantic alignment with user needs.

Although TF3 (Lack of Contextual Understanding) cannot be adjusted directly by educators or students, it is essential to include it in the model because it defines a baseline limitation of current GenAI tools—one that educators must work around and students must learn to recognize. Including this factor helps explain why students must be trained to critically assess GenAI outputs rather than trust them blindly. It also justifies the importance of professor-led interventions such as PF3 (In-Class Demos of GenAI Limits) and PF7 (AI Output Validation Tasks), which are designed precisely to mitigate the effects of TF3. In this way, TF3 functions as a structural constraint that informs the design of responses elsewhere in the system. Ignoring it would leave a critical blind spot in understanding how misuse or misunderstanding of GenAI arises—even when all other pedagogical factors are well implemented.

The fourth GenAI-tool-related factor (TF4) is about the AI hallucination rate, another widely discussed issue of AI [56-58]. It is the tendency of GenAI tools to produce seemingly valid, but logically invalid, or fabricated outputs, which would pose a direct risk to students' learning. Students unfamiliar with this behavior may take such outputs at face value, particularly if GenAI tools present the outputs with fluency and confidence.

Together, these four tool-level factors underscore the importance of embedding AI literacy and validation behaviors into SEEd pedagogy. They also highlight that responsible GenAI use is not only a matter of student intent or instructional policy, but also of how the tools themselves are designed and perceived during use.

#### 3.1.4 Educational outcome factors

The model considers eight educational outcomes, labeled with terms starting with "O" in the model (Figure 2). These outcomes serve as cognitive, behavioral, and ethical goals that student–tool–instructor interactions should support.

First, the model emphasizes using AI's support for conceptual understanding of SE topics (O1). GenAI tools, when used thoughtfully, can help clarify complex technical content and support novice learning. However, this benefit must be balanced by resisting misuse of AI in core SE tasks (O2), such as design, testing, and debugging—areas where reliance on AI can undermine deep learning if used inappropriately.

Students are also expected to progress toward developing AI-assisted SE skills (O3), which involve integrating GenAI into their engineering workflows in ways aligned with real-world practices. A critical component of this process is the reflection and critical evaluation of AI artifacts (O4), such as examining the structure, assumptions, or limitations of generated code, documentation, or diagrams—reflecting higher-order cognitive engagement as defined in Bloom's taxonomy.

Additionally, students must focus on developing and preserving critical thinking skills during AI-supported learning (O5). As GenAI lowers technical barriers, students must still be trained to engage analytically, ask probing questions, and weigh alternatives. This is complemented by developing skills to verify AI-generated outputs (O6)—a distinct outcome focused on oversight, accuracy, and the systematic detection of errors or hallucinations in AI responses.

While factors O6 and O4 are related, they are not the same, as they target different cognitive processes and learning goals. While O6 emphasizes factual correctness and functional validation, O4—reflection and critical evaluation of AI artifacts—focuses on the conceptual and strategic appraisal of AI outputs, such as assessing their appropriateness, design rationale, or alignment with task goals. Together, these outcomes target both technical accuracy and higher-order reasoning, reinforcing comprehensive AI literacy. An example of O4 could be: Is the design pattern, provided by the AI, appropriate for the use case? An example of O6 could be: Does the test case provided by the AI, actually cover the edge condition?

To ensure ethical practice, the model includes upholding academic integrity in AI-supported learning environments (O7). As GenAI becomes more accessible, students must understand boundaries, attribution expectations, and responsible use. Lastly, the model highlights building confidence in judging and using AI responsibly (O8), helping students not just adopt tools, but develop the judgment to know when and how to use them—and when not to.

Together, these outcomes represent a comprehensive vision of what it means to use GenAI responsibly in SEEd, blending skill development, ethical behavior, critical reflection, and learner autonomy.

## 3.2 Structure and relationships of the explanatory model

The model consists of four interconnected components: Professor Factors (PF1–PF10), Student Factors (SF1–SF8), Tool Factors (TF1–TF4), and Educational Outcomes (O1–O8). Together, these elements form a sociotechnical system that captures the dynamics of GenAI use in SEEd. The model reflects both human and technical dimensions—ranging from pedagogical design and student dispositions to the functional behaviors of GenAI tools and the nature of SE artifacts.

We emphasize that SEEd differs from many other disciplines in the types of GenAI-generated outputs students are expected to engage with. While students in the humanities may use GenAI to produce text-based artifacts such as essays or summaries, SEEd involves the generation of structured, executable, and domain-sensitive artifacts—including source code, test cases, bug reports, requirement specifications, and architectural diagrams. These artifacts are not only syntactically complex but also require logical soundness, design consistency, and alignment with engineering principles. As such, effective use of GenAI in SEEd demands not just productivity but critical oversight and domain-specific validation.

This distinction has important implications for the structure of our model as we discuss next.

<u>Professor-related Factors:</u>

Professor-related Factors (PF), for example, influence not only general GenAI behavior but also students' approach to interacting with SE artifacts. For instance, assignment design (PF1) and validation tasks (PF7) must be tailored to teach students how to scrutinize AI-generated code or test cases for correctness, coverage, and maintainability. Instructors who model AI usage (PF5) by walking through flawed GenAI-generated artifacts help students build habits of reflective and technical critique—directly supporting outcomes like O4 (Reflection and critical evaluation of AI artifacts) and O6 (Verification skills). The degree to which GenAI is formally integrated into the course (PF10) also affects whether students perceive these tools as exploratory aids or as authoritative sources—shaping their attitude toward output validation.

<u>Student-related Factors:</u>

Student-related Factors (SF) such as AI literacy (SF1), time pressure (SF2), and prior AI-tool experience (SF5) interact heavily with the demands of SE artifacts. Students under time constraints may rely on GenAI to generate complete scripts or test cases without reviewing logic or edge conditions. Similarly, those with limited experience in debugging or design may find AI outputs deceptively fluent—leading to misuse (O2) or failure to verify (O6). Importantly, peer norms (SF4) around "acceptable" levels of GenAI use for artifact generation can normalize either healthy or harmful practices.

<u>Tool-related Factors:</u>

Tool-related Factors (TF)—including ease of use (TF1), output quality (TF2), lack of contextual understanding (TF3), and hallucination rate (TF4)—take on heightened significance in SEEd. These tools often produce code that compiles but fails semantically, or design diagrams that appear structurally valid but violate basic architectural principles. Students must therefore not only engage with the visible structure of artifacts, but also evaluate their semantic integrity—an expectation embedded in outcomes such as O3 (Developing AI-assisted SE skills), O5 (Critical thinking), and O6 (Oversight).

<u>Educational Outcomes:</u>

Educational Outcomes (O1–O8) articulate the skills and dispositions that responsible GenAI use in SEEd should foster. The model makes clear distinctions: for example, O6 (Developing skills to verify AI-generated outputs) targets technical correctness, while O4 (Reflection and critical evaluation) emphasizes strategic and conceptual reasoning about SE artifacts. O1 (Using AI support for conceptual understanding) is particularly relevant when GenAI is used to generate simplified code examples or explain complex testing concepts. O3 (Developing AI-assisted SE skills) represents a core translational goal: enabling students to integrate GenAI into professional SE workflows without overreliance or ethical lapses.

The educational outcome O6 is of special importance, since students need to develop skills to verify AI-generated outputs. This outcome aims to help students build the ability to critically assess and validate artifacts produced by GenAI tools. Rather than accepting AI-generated content at face value, students shall be encouraged to treat it as a draft—subject to inspection, adaptation, or rejection. In this context, verification tasks include checking factual accuracy, alignment with assignment requirements, and consistency with SE best practices. Students shall also be asked to identify and correct common issues in GenAI outputs, such as hallucinations, vague statements, or logical flaws.

The verification process can be informal and instructor-led. The assignments and review checkpoints shall encourage reflection and correction, reinforcing the need for thoughtful engagement with GenAI tools. Importantly, this verification mindset shall also strengthen students' peer-review and code inspection skills. By analyzing both accurate and flawed AI-generated outputs, students need to develop sharper judgment and constructive feedback capabilities—skills essential for effective SE practice. It is also recommended to introduce students to latest developments in the area of verification and quality assessment of AI-generated SE artifacts, e.g., [59, 60].

Finally, the model includes feedback loops. For example, students who gain confidence (O8) in using GenAI responsibly may improve in verification (O6) and reflection (O4) in future tasks. Likewise, instructor policies and practices may evolve based on observed misuse patterns or deficiencies in output validation—thus completing a pedagogical feedback cycle.

Summary of the model:

In summary, the model reflects the discipline-specific realities of SEEd, where GenAI is used not to generate prose but to produce technically meaningful artifacts. These artifacts—by their nature—require what we term *layered interpretation*: students must evaluate not only surface correctness (e.g., does the code compile?) but also deeper qualities such as logical validity, coverage of edge cases, alignment with SE design principles, and even attribution ethics. The model provides a framework for aligning instructional strategies, tool properties, and student factors to ensure that GenAI use fosters authentic, rigorous, and industry-relevant learning outcomes in an AI-enhanced SE education landscape.

### 3.3 Validating the model's internal consistency

When developing the proposed model, it was important to validate the model's internal consistency [61]—that is, the degree to which its constructs (the different factors: PFs, SFs, TFs, and Os) are conceptually coherent, mutually reinforcing, and logically interconnected. We validated this consistency through theoretical grounding, systematic internal analysis, and preliminary external feedback.

First, we ensured that the model's design is aligned with well-established frameworks in educational research, including constructivist learning theory [44], sociotechnical systems thinking [45], and Bloom's taxonomy [31, 33, 43]. In the internal consistency validation process, we observed that Professor factors (PFs) reflect evidence-based pedagogical levers such as assignment design, modeling, and feedback. Student factors (SFs) capture motivational, cognitive, and social influences on GenAI behavior. Tool factors (TFs), while not modifiable by educators, represent structural affordances and constraints that directly shape student interactions with AI.

Each component maps logically to one or more targeted educational outcomes (Os), with attention to minimizing overlap or ambiguity. For instance, while O4 (Reflection and critical evaluation) and O6 (Verification of AI outputs) may seem related, the former emphasizes conceptual judgment, whereas the latter involves technical correctness—an important distinction in SEEd contexts.

To ensure internal logic, we examined the causal relationships across the model. Each link was reviewed to confirm that it represented a plausible, literature-informed relationship. For example, PF7 (AI output validation tasks) clearly supports O6 (Verification), while TF4 (Hallucination rate) poses a threat to that same outcome if left unaddressed. Feedback loops were also accounted for: O8 (Confidence in judging AI responsibly), once achieved, may reinforce behaviors that support other outcomes like O2 (Resisting misuse) and O5 (Critical thinking).

Preliminary feedback was also obtained from instructors at two institutions (details in Section 4) who reviewed the model's constructs and causal logic. Their feedback confirmed that all the factors were recognizable and actionable in their own teaching contexts. While this initial validation was informal, it reinforced the model's pedagogical plausibility and classroom relevance.

To further ensure coherence, we cross-referenced our model’s constructs with established frameworks such as Bloom’s taxonomy AI-aware version [43] and UNESCO’s AI literacy guidelines [11, 12], ensuring conceptual alignment with broader educational discourse. This comparison validated that our outcome factors map well onto widely recognized cognitive and ethical learning goals, particularly in the context of AI-supported education.

As we will see in Section 4, the model was applied in course planning scenarios, where its categories and relationships provided interpretable, non-contradictory guidance. This practical application affirmed not only its theoretical soundness but also its utility as a tool for curriculum design, instructional decision-making, and pedagogical reflection in SEEd environments.

While the current validation has been primarily qualitative and based on instructor feedback, future work will include more structured evaluation of the model's validity. Planned studies will involve quantitative methods such as rubric-aligned assessment of student artifacts, pre/post surveys on GenAI use and critical thinking, and expert reviews of construct definitions. We also aim to conduct focus groups with students and industry mentors to refine the model’s terminology, assumptions, and practical relevance.

### 3.4 Mapping factors to intervention design

To support real-world application, each factor in the model can be linked to one or more instructional interventions, enabling educators to address specific challenges associated with GenAI use in SEEd. This mapping serves as a practical bridge between theory and classroom action, guiding instructors in transforming the model’s constructs into course-level strategies.

For instance, factors related to professor behavior—such as modeling proper GenAI use (PF5), providing AI output validation prompts (PF7), or offering in-class demonstrations of GenAI’s limitations (PF3)—can be directly operationalized within assignment scaffolding, rubric criteria, or lecture activities. These interventions aim to help students build skills in verification (O6), reflection (O4), and responsible use (O8). For example, an instructor might walk through a flawed AI-generated test case during class, highlighting subtle logic errors to promote student metacognition and skepticism.

Factors relating to student-side influences, such as peer pressure (SF4) or perceived risk of getting caught (SF8), can inform course-level policy and culture. These could be addressed via transparent AI usage policies, structured classroom discussions about ethical use, or reflective journaling where students document and justify their GenAI engagement. Similarly, student motivation type (SF3) and AI literacy (SF1) can be shaped through rubric-aligned expectations, tutorials on the limitations and failure modes of GenAI tools, or self-assessment activities where students reflect on their own learning vs. AI reliance.

Tool-side factors also require pedagogical attention, even if they are not directly controllable by instructors. For example, output fluency (TF2) and hallucination rate (TF4) can be turned into learning opportunities by assigning students to critique, debug, or cross-validate GenAI-generated artifacts. Instructors might create activities where students compare human-written and AI-written bug reports or ask students to refactor GenAI-generated code for clarity, performance, or correctness—thereby turning tool limitations into active learning opportunities.

This mapping is not rigid; rather, it offers a flexible guide for instructors to align learning outcomes with concrete practices, tailored to course level, student demographics, and institutional context. The interventions can range from minor changes (e.g., adding one rubric criterion) to structural redesigns (e.g., integrating AI-reflection assignments across a term). What matters is that the mapping helps activate the model’s insights into targeted design decisions—enabling instructors to scaffold not just assignments, but student thinking, habits, and ethical stances.

By making explicit the links between factors and tangible pedagogical interventions, this section positions the model as a practical tool for instructional design.

### 3.5 SEEd-specific features of the explanatory model and its adaptability to other disciplines

While the model presented in this study was designed for SEEd, many of its components—such as professor modeling, AI policy clarity, and student validation behaviors—can be applied more broadly across educational domains. That said, the model intentionally incorporates features that reflect the distinct nature of SEEd. For example, the types of GenAI-generated artifacts in SEEd (e.g., source code, test plans, debugging instructions, or UML diagrams) are structurally different and more logic-dependent than those in fields like humanities or social sciences, where outputs often include essays, summaries, or discussions. Validating such structured artifacts requires students to engage with syntax correctness, logical flow, test adequacy, and functional behavior—skills fundamental to SE practice.

Moreover, SE tasks often demand rigorous reasoning, systematic debugging, and iterative improvement—competencies that may be bypassed if students uncritically accept AI outputs. The model therefore prioritizes student behaviors such as peer validation of AI-generated code and reflective evaluation of outputs, and it positions the instructor not only as a rule-setter but as an AI-usage role model.

Despite this domain grounding, the model is transferable and adaptable to educational contexts of other fields. Fields involving structured problem solving—such as data science, electrical engineering, or economics—may similarly benefit from explicit validation tasks and self-reflections on AI usage. As such, while this study focuses on SEEd, the model offers a flexible framework for encouraging responsible GenAI use across many disciplines.

### 3.6 Comparison with other related models / frameworks of responsible AI use in education

In developing our model, we reviewed and considered several existing frameworks that address responsible AI use in educational contexts (as briefly summarized in Section 2.5). While those models offer important contributions, our approach differs in both structure and scope, particularly in its focus on causal relationships, SEEd-specific artifacts, and practitioner-centered interventions. Below, we compare our model with five prominent frameworks:

- The two AI-in-education competency frameworks by UNESCO, one for students [11] and the other for teachers [12]
- The PAIR framework (Problem, AI, Interaction, Reflection) [36]
- The AI-ICE model (Interaction, Cognition, Ethics) [37]
- The AI-Lab framework [38]

The UNESCO AI Competency Frameworks for students [11] and teachers [12] provide insightful high-level guidelines for developing AI-related knowledge, skills, and ethical awareness. These frameworks emphasize AI literacy, digital skills, and responsible participation in AI-mediated environments. However, they are structured as competency checklists rather than causal models, and they are not tailored to the unique demands of SEEd. In contrast, our model focuses on how specific teaching practices, student dispositions, and tool characteristics dynamically interact to shape outcomes like verification (O6) or reflection (O4), particularly for specific SE artifacts such as code and test cases.

The PAIR framework (Problem, AI, Interaction, Reflection) by [36] offers a pedagogically grounded 4-phase approach to integrating GenAI into student learning workflows. While PAIR usefully emphasizes the importance of contextual prompts and student reflection, it functions more as a lesson-level design guide than a systems-level model. Our model builds on this insight by specifying *what types of professor actions and student traits* (e.g., PF5, SF4) most influence such reflection, and how they map to diverse educational outcomes.

Similarly, the AI-ICE model (Interaction, Cognition, Ethics) [37] categorizes GenAI's educational impacts into three domains: how students interact with AI, how it affects cognition, and its ethical dimensions. While this tripartite structure is useful for framing broad concerns, it lacks granularity and actionable instructional guidance. Our model complements AI-ICE by offering more concrete constructs (e.g., PF7: validation tasks, SF1: AI literacy) and specifying causal links to outcomes like critical thinking (O5) or academic integrity (O7).

The other framework, AI-Lab [38], is more closely aligned with SEEd, as it provides support for integrating GenAI tools into programming courses. It emphasizes iterative experimentation and reflective practices. However, AI-Lab functions primarily as a course-structure template, not as an explanatory or predictive model. Our model extends this work by offering a theory-driven map of *why* certain interventions succeed or fail, especially in managing SE-specific artifacts where hallucination or superficial correctness can mislead students.

Unlike the UNESCO frameworks or PAIR/AI-ICE models, which focus primarily on broad competency development and ethical awareness, our model explicitly maps causal relationships between educator behavior, student conditions, and observable outcomes in GenAI use. This structure makes it suitable not only for curriculum alignment but also for course-level intervention design and evaluation. Moreover, our emphasis on SEEd-specific artifacts (e.g., test cases, architectural diagrams) distinguishes the model's practical utility in technical computing contexts.

In summary, while all the above models have contributed valuable perspectives, our model is unique in four respects:

1. It provides a multi-actor causal structure, linking professor, student, and tool factors to targeted outcomes;
2. It is SEEd-specific, grounded in the unique properties of software artifacts (code, tests, diagrams);

3. It emphasizes pedagogical interventions, not just competencies or workflows; and
4. It was refined through empirical feedback and alignment with existing educational theory (e.g., Bloom's Taxonomy, and UNESCO AI guidelines).

By positioning our model w.r.t those related works, our model aims to advance the field with a flexible, actionable framework tailored to the realities of GenAI use in SEEd and other fields.

## 4 EMPIRICAL APPLICATIONS AND ASSESSMENTS OF THE APPROACH

We have initiated the application and assessment of our proposed approach in two different educational contexts: (1) using the approach for improving a final-year software testing course, by making it GenAI-aware; and (2) using the approach in the redesign exercise of an undergraduate SE program. We discuss each empirical application and the observations in the following sections.

To provide a structured evaluation of the explanatory model, we report its application across two settings following a common format: (1) context, (2) design of intervention, (3) data sources, and (4) observations aligned with model factors. This structure ensures consistency and clarity in how evidence from each case is presented.

### 4.1 Empirical application #1: Using the approach for making a final-year software testing course AI-aware (TBD)

TBD

### 4.2 Empirical application #2: Using the approach in the design of an undergraduate software engineering program

As the second empirical setting, we applied the model during the redesign process of an undergraduate SE degree program. Unlike the first empirical application (Section 4.1), which involved revising an existing course, this setting allowed us to integrate GenAI-related competencies, responsibilities, and validation practices into the foundational structure of a new SE program. This section describes the context, intervention strategy, and early reflections from that degree's curriculum design process.

#### 4.2.1 Context: Designing a Software Engineering undergraduate program

The academic department in which this program is offered, has recently decided to the initiative of designing a new undergraduate SE degree program. Prior to this, SE was typically taught as a specialization within broader IT or Computer Engineering programs. This new program will be a four-year, BSc program focused on the theory and practice of SE.

The program's design team included SE researchers, academic leaders, and industry consultants, and emphasized modernity, relevance, and international alignment. Among the paper's co-authors were the participants in the program's design team. The curriculum integrates classical SE topics such as requirements, architecture, testing, maintenance, and project management, with emerging and industry-critical areas such as DevOps, cybersecurity, and AI-assisted SE.

From the outset, the design team recognized the importance of equipping graduates not only with technical skills but also with the ability to use GenAI tools responsibly. This was especially relevant considering the growing GenAI adoption across global software industries. Inspired by our model of responsible GenAI use in SE education (Section 3), the team sought to embed model-aligned design principles directly into the curricular structure. This ensured that responsible GenAI engagement would be supported longitudinally in the entirety of the new 4-year-long SE program, not relegated to one-off discussions or isolated courses on AI ethics.

Our design-based intervention took place during the program development process and was guided directly by our model. The objective was to systematically embed GenAI awareness and responsibility across the curriculum in a way that balanced flexibility with pedagogical integrity.

#### 4.2.2 Design of intervention

##### 4.2.2.1 Curriculum-level GenAI integration

Several mid- and upper-year courses were selected for targeted GenAI-aware enhancements: Software Testing, Software Architecture, and Software Cybersecurity courses were updated to include learning outcomes related to GenAI artifact critique, validation, and trust.

Assignments in these courses now require students to disclose GenAI use, perform critical reviews of GenAI-generated test cases, architecture diagrams, or threat models, and justify whether the outputs were accurate and context-appropriate. Module rubrics were refined to reward validation and oversight—rather than penalize GenAI use outright. This operationalizes PF2 (usage policy), PF7 (validation tasks), and O6 (AI output verification) from the model.

##### 4.2.2.2 Program-wide AI-literacy

To support sustainable adoption of AI in the SE program, we applied a longitudinal pattern as discussed next. In early-semester courses (e.g., Introduction to Programming, Data Structures), instructors introduce basic GenAI concepts, enabling students to reflect on when and why to use such tools. In core SE courses (e.g., SE Principles), students were given practical exercises that ask them to compare human-generated vs. GenAI-generated content (e.g., requirements, UML diagrams).

In later-stage courses, students were expected to critically validate AI-generated artifacts, including through group work and capstone projects, using instructor-designed peer review or critique rubrics. In the final-year (capstone) project course, students must submit a brief appendix indicating if and how GenAI tools supported their technical development work—and what steps were taken to verify those outputs.

This multistage approach corresponds with a cognitive and motivational progression across the program, ensuring students move from basic AI awareness (SF1) to higher-order reasoning (O4, O6), while also reinforcing career-oriented messaging (PF6, O8).

##### 4.2.2.3 Institutional alignment

To support the intervention beyond individual courses of the new SE program, several institutional-level efforts are being initiated to ensure consistency and long-term cultural change in how GenAI is approached across the computing curriculum.

Faculty development sessions are being proposed to help instructors navigate the pedagogical and ethical complexities of GenAI use. These sessions aim to provide practical strategies for designing assignments that promote validation, reflection, and responsible integration of GenAI tools. Instructors will also be supported in developing grading approaches that reward critical oversight rather than surface-level output, directly reinforcing outcomes such as O4 (Reflection) and O6 (Verification).

Course templates and assignment guidelines are being revised to include model-informed language regarding GenAI use. This includes clarifying expectations for citation, validation, and acceptable tool integration, as well as embedding outcome-aligned AI prompts. By standardizing how GenAI is addressed in course materials, the institution aims to reduce student confusion and promote consistent learning messages across modules—supporting O7 (Academic Integrity) and PF9 (Policy Clarity).

Career services are actively exploring ways to align with the model's emphasis on real-world AI readiness. They are developing messaging and resources to help students articulate their GenAI usage practices in interviews and professional settings. This includes highlighting the importance of explaining validation steps, ethical considerations, and limitations of AI-generated artifacts—reinforcing model outcomes such as O3 (AI-SE Skills) and O8 (Confidence in Judging and Using AI Responsibly).

Together, these institutional measures aim to create a supportive ecosystem in which responsible GenAI use is coherently promoted, assessed, and rewarded—not just within isolated courses, but across the educational journey.

##### 4.2.2.4 Mapping of AI-related Program-level Outcomes to SE courses, their assessments and rubrics

Explicitly mapping program-level outcomes to courses, assessments, and rubrics would help demonstrate how institutional goals are operationalized in practice and how students' attainment of AI-related SE skills is systematically evaluated. We indeed observed this need in our empirical application #2.

Table 2 illustrates how program-level outcomes related to AI-based SE skills can be explicitly mapped to several key SE courses, their assessments, and the corresponding rubrics for grading assessments within a typical SE program. The five outcomes—responsible application of AI, inspection of AI-generated artifacts, integration into SE processes, evaluation of risks and ethics, and collaborative use of AI tools—span multiple courses across the program. Each course includes assessments where students engage with AI in context, and the associated rubrics ensure that performance is measured not only on technical correctness but also on responsible and critical use of AI.

**Table 2- Mapping of AI-related Program-level Outcomes (PLOs) to several key SE courses, their assessments and rubrics**

| Program-Level Outcome (PLO) | Courses | Assessments | Rubric Element |
|---|---|---|---|
| Apply generative AI responsibly across SE activities | Intro to SE, Software Project Management | Short essays on AI use cases; project planning assignment | Clarity in describing appropriate AI use and limits |
| Critically inspect and review AI-generated artifacts | Requirements Engineering, Software Testing | Requirements specs with AI assistance; lab reports | Accuracy in identifying AI errors and proposing corrections |
| Integrate AI outputs into SE processes | Software Design, Software Architecture | Design diagrams/code with AI suggestions; architecture assignment | Justification of integration choices and validation steps |
| Evaluate risks and ethics of AI in SE practice | Professional Practice, Software Quality Assurance | Case study analysis; QA plan including AI considerations | Demonstration of risk awareness and ethical reasoning |
| Collaborate using AI-enabled SE tools | Capstone Project, DevOps & CI/CD | Team project deliverables; pipeline setup with AI-assisted tools | Effective team integration of AI tools and transparent reporting |

### 4.2.3 Data sources

Evidence for this empirical application was derived from multiple institutional artifacts. First, curriculum design documents and draft syllabi provided the foundation for mapping program-level outcomes (PLOs) to courses and assessments. Second, records from program design workshops and discussions with faculty offered insights into how GenAI-related skills could be embedded at scale. Third, the draft mapping tables—linking PLOs to specific courses, assignments, and rubrics—served as a concrete basis for evaluating how the explanatory model informed program-level design choices. Together, these sources created a structured dataset for reflecting on the applicability of the model beyond a single course context.

### 4.2.4 Reflections and observations

Several early insights have emerged from the design and stakeholder validation process, as we discuss next. Faculty, curriculum designers, and reviewers from industry and the Ministry of Education agreed that GenAI awareness should not be siloed into ethics modules, but rather infused across the entire program. The model helped make this integration tangible.

Some early reviewers questioned if first-year students could meaningfully reflect on GenAI use. However, the model's actor-based and outcome-based design allowed us to scale GenAI expectations to student readiness: from exposure and transparency in Year 1, to full artifact critique and validation in Years 3 and 4.

The model has supported both vertical and horizontal coherence. It has helped the design team balance learning progression across semesters, while ensuring that assignments in different courses (e.g., Software Architecture and Testing) reinforced overlapping responsible use outcomes (e.g., O4 and O6).

Curriculum-level integration has allowed the program to model GenAI literacy as a professional SE competency, not an optional skill. This was well received by industry partners who emphasized that overreliance on GenAI without critical review was already a problem in junior developer hiring pipelines.

These reflections confirm that the model can function not only as a micro-level course redesign tool but also as a macro-level curriculum design framework, capable of supporting SE education programs that take GenAI seriously as both a pedagogical opportunity and a professional challenge.

## 5 REFLECTIONS AND LESSONS LEARNED, IMPLICATIONS AND LIMITATIONS

We now reflect on our experience applying the model of responsible GenAI use in SEEd in two distinct educational contexts: a final-year Software Testing course and a national-level undergraduate curriculum design effort. We summarize key lessons learned, outline practical implications for educators and curriculum designers, and critically examine the model's current limitations. These reflections aim to inform others seeking to embed GenAI literacy and responsibility into SE and broader computing education.

## 5.1 Reflections and lessons learned

Our experience applying the model across both a course (Section 4.1) and the design of a full new SE program (Section 4.2) revealed that GenAI-related pedagogical challenges cannot be effectively addressed through isolated policies or reactionary controls. Instead, meaningful change requires targeted, theory-informed design interventions that align with instructors' goals, student realities, and AI tools' characteristics.

At the course level (Section 4.1), we found that even lightweight changes—such as adding usage declarations, validation prompts, and career warnings—can shift how students think about and engage with GenAI tools. Instructor reflections and informal student responses suggest that these revisions supported increased awareness, critical evaluation, and reduced passive reliance on GenAI.

At the program level (Section 4.2), the model provided a shared vocabulary and structure that enabled coordinated, longitudinal planning. It helped us see where GenAI literacy could be scaffolded over time, and how learning outcomes in various modules could align with a coherent, developmental trajectory of responsible AI use.

The model also proved valuable as a tool for real-time pedagogical decision-making during implementation. Instructors—and, in some cases, teaching assistants or curriculum reviewers—used the model to navigate uncertainty, such as whether to include GenAI prompts in first-year labs or final-year capstone projects. By identifying which outcomes (e.g., O1 or O8) and influencing factors (e.g., PF1, SF1) were most relevant, the model helped clarify the educational goals at stake and informed principled trade-off analysis.

A key reflection from this work is that the explanatory model has already undergone two design-based research (DBR) iterations. The first cycle in the software testing course allowed us to observe student behavior directly and test the usefulness of interventions such as inspection prompts and usage declarations. The second cycle, at the program level, demonstrated how the model can guide institutional mapping of program learning outcomes to assessments and rubrics. These iterations highlight that the model is not static but evolving, and each application provides lessons that inform its refinement and adaptation to new contexts.

## 5.2 Implications for educators

This work offers several actionable insights for SE and CS educators seeking to respond constructively to the rise of GenAI tools in student learning environments. Rather than banning or ignoring these tools, our model supports a **responsibility-centered approach** that builds students' critical thinking and professional readiness.

We offer the following practical suggestions:

- **Encourage validation and critique of GenAI outputs**: Rather than prohibit tool usage, instructors can require students to verify or critique the AI-generated artifacts they submit—such as test cases, bug reports, or design diagrams. Prompts like "How did you verify this output?" or "What are its limitations?" help cultivate oversight and reinforce learning outcomes like O4 (reflection) and O6 (verification).
- **Leverage peer review to broaden accountability**: Including peer-review activities that focus on GenAI usage not only encourages deeper engagement by the original author but also builds shared responsibility and critical assessment skills among reviewers. Students begin to develop a more nuanced understanding of what constitutes responsible use.
- **Integrate career-relevant prompts into coursework**: Framing assignments with references to workplace expectations—e.g., "How would you explain this AI-generated report in a job interview?"—helps students see the connection between responsible GenAI use and future performance. This approach supports outcomes like O3 (AI-SE skills) and O8 (confidence in using AI responsibly).

Beyond these course-level strategies, our model has broader generalizability to computing education. While our work focused on SEEd, many elements—such as actor groups (students, instructors), tool factors (e.g., output quality, hallucination risk), and outcome constructs (e.g., reflection, oversight)—are applicable across computing domains. In areas like systems programming, databases, HCI, or cybersecurity, GenAI tools are increasingly used for generating code, writing

SQL queries, modeling scenarios, or producing reports. The same principles of critical engagement, output validation, and ethical integration apply, with minimal adaptation.

Finally, our experience highlights the value of using explanatory models and design-based research (DBR) approaches in AI-era curriculum innovation. DBR enabled iterative refinement through real classroom feedback, while the causal model provided a structured framework for reasoning about the relationships among instructor practices, student behaviors, tool characteristics, and learning outcomes. Together, they form a coherent toolkit for educators—not just to respond reactively to GenAI, but to proactively design learning environments that promote responsible, reflective, and career-aligned AI use.

In summary, educators across computing disciplines can adapt this model to design interventions that align with their specific contexts and learning goals, helping students move from passive AI consumers to active, critical users equipped for real-world SE and beyond.

## 5.3 Limitations of the model and threats to validity

While our work offers a novel and actionable approach to promoting responsible GenAI use in SEEd, it has several limitations that merit discussion. We discuss the limitations in two parts: (1) limitations of the model; and (2) limitations (threats to validity) of the empirical applications

### 5.3.1 Limitations of the model

Although our explanatory model provides a structured lens to analyze and guide responsible GenAI use in SEEd, several limitations must be acknowledged.

First, the model was developed primarily through design-based research in SE contexts. While it integrates concepts from broader educational theory (e.g., Bloom's taxonomy, constructivism, sociotechnical systems), its constructs may not fully capture nuances in other domains or institutional settings.

Second, the model emphasizes causal relationships among professor, student, and tool factors, but these relationships remain theoretical until further validated through larger-scale, multi-institutional studies. Third, although we distinguished overlapping constructs such as reflection (O4) and verification (O6), conceptual overlaps may still emerge in practice, and more refined operational definitions may be necessary.

Fourth, the model currently assumes a relatively stable tool landscape; however, the rapid evolution of GenAI technologies may introduce new factors (e.g., multimodal outputs, adaptive personalization) that require extensions to the framework.

Finally, while the model incorporates both educationally-relevant outcomes and discipline-specific features of SEEd, it does not address technical SE performance metrics (e.g., test coverage, defect detection), as these were outside the educational focus of this work.

### 5.3.2 Threats to validity of the empirical applications of the model

The two empirical applications of the model—course-level intervention in a Software Testing course and program-level curriculum design—also carry limitations and threats to validity. We discuss these in terms of the four common categories of validity threats [62]: internal validity, construct validity, conclusion validity and external validity, and also the steps that we have taken to minimize or mitigate them.

**Internal validity.** In the course-level study, student behavior changes may have been influenced by factors beyond the interventions, such as workload distribution or the issue of other parallel courses. Similarly, at the program level, the design process was shaped by institutional constraints and stakeholder priorities, not solely by our model.

**Construct validity.** Although we carefully aligned survey items, rubrics, and reflections with the model's outcomes, constructs such as "critical reflection" or "responsible use" remain partly subjective. The interpretation of these constructs by students and instructors may not fully match our intended definitions.

**Conclusion validity.** The sample sizes were modest and limited to single institutions, which restricts the generalizability of statistical findings. The positive survey results indicate promising trends, but they should be interpreted cautiously given possible novelty effects of AI-focused interventions.

**External validity.** Both case studies were conducted in specific cultural and institutional contexts (Azerbaijan and the UK). While the model has broad conceptual grounding, its transferability to other settings requires further replication.

Curriculum design choices, in particular, may vary significantly depending on national accreditation standards and institutional readiness.

Additionally, selection bias cannot be ruled out. Students who engaged more actively with GenAI tasks may have been predisposed to experiment with new tools, whereas disengaged students may have underreported or superficially complied. Novelty effects may also have amplified positive responses, as AI-focused interventions were new and potentially perceived as innovative.

Together, these limitations underscore that our results should be treated as initial evidence rather than definitive validation. Broader replication, triangulation with additional data sources, and longer-term follow-up are needed to establish stronger validity.

## 6 CONCLUSIONS AND FUTURE WORKS

This paper presented a principled and empirically grounded approach to promoting responsible use of GenAI tools in SEEd. We proposed an explanatory model that outlines the factors influencing student behavior in GenAI-assisted learning and used it to guide the redesign of lab activities in an undergraduate SE testing course. Our goal was not to restrict GenAI usage, but to build critical thinking, verification, and reflection into AI-supported workflows.

The lab design was carefully mapped to specific elements of the model, enabling targeted interventions. These included encouraging context-aware prompt formulation, requiring students to inspect and revise AI-generated artifacts, and incorporating peer review milestones. The model not only supported the instructional design but also helped us evaluate the effectiveness of the interventions.

Technical analysis of student outputs showed that while GenAI tools helped students complete tasks faster, they often introduced issues such as hallucinated content, vague logic, and redundant test cases. Students learned to take a more critical stance—inspecting outputs carefully and correcting them when necessary. This helped reinforce core SE skills related to quality assurance, logic, and precision.

Prompt engineering emerged as an important skill. Students found that the quality of GenAI outputs was directly tied to the clarity and specificity of their prompts. Through iterative refinement, they improved their instructions and developed a better understanding of how to guide GenAI tools effectively. This highlights a shift in required competencies—from simply producing artifacts to strategically steering GenAI toward useful outputs.

Comparing GenAI-generated outputs with manually created artifacts from previous course offerings was particularly valuable. These comparisons helped students understand what makes an output correct, complete, or relevant. This process made verification more concrete, grounded in actual examples rather than abstract expectations. It also gave students a clearer sense of how much human judgment is still required when using AI in SE contexts.

Overall, the model proved helpful not only in guiding lab design but also in interpreting how students engaged with GenAI tools. It provides a practical framework that other educators can apply to balance the opportunities and risks of GenAI integration. We believe this model strikes a middle ground—embracing innovation while maintaining the rigor needed for meaningful learning in SE education.

For future work, we plan to apply this model to other areas of SE education such as software design, documentation, and maintenance. We also aim to gather more quantitative data to measure learning outcomes and student behavior.

While this paper concentrates on educationally-relevant objective measures to evaluate students' responsible use of AI, future work could complement these by incorporating more technical software testing metrics. Examples include code coverage, defect detection rates, or structured scoring of AI-artifact inspections. Such measures would extend the current educational focus with a technical perspective, enabling a more holistic evaluation that combines learning outcomes with testing effectiveness.

Also, we hope the model contributes to broader conversations about responsible GenAI use in computer science curricula and supports the development of policy and pedagogy aligned with real-world SE practices.

## ACKNOWLEDGEMENTS

The authors note that they have shared a pre-print of this work on the arXiv platform [63] for well-established purposes: to enable early dissemination, accelerate scientific communication, receive early feedback from peers, increase the work's visibility, provide a citable reference, and foster collaboration and dialogue with other researchers.

## Appendix A- The checklist for inspection of AI-generated outputs, that was provided to students

Students were asked to conduct validation, in the form of inspection of AI-generated artifacts. This activity is also called human oversight of AI. Students were asked to use the following inspection checklist. See Section 4.1.2 for more details.

| # | Checklist item | Scale |
|---|---|---|
| 1 | The artifact is relevant to the task and meets the lab requirements. | 1–5 Likert scale (Strongly disagree to Strongly agree) |
| 2 | The artifact is logically coherent and free of contradictions. | 1–5 Likert scale |
| 3 | The artifact contains appropriate technical terminology and is understandable. | 1–5 Likert scale |
| 4 | There are no signs of hallucination or fabricated content. | Yes / No |
| 5 | The artifact follows a structure similar to examples taught in class. | 1–5 Likert scale |
| 6 | The artifact includes necessary details (e.g., preconditions, steps, expected outcomes). | 1–5 Likert scale |
| 7 | The artifact shows signs of generic or vague content that may not reflect real system behavior. | Yes / No |
| 8 | You were able to trace how the AI generated the artifact based on the given prompt. | Yes / No |
| 9 | You identified and corrected at least one issue in the AI-generated artifact during your review. | Yes / No |
| 10 | The student recognizes that the quality of the AI-generated artifact depends on the prompt provided, and believes that revising the prompt could lead to a better result. | Yes / No |